\documentclass[preprint,showpacs,preprintnumbers,amsmath,amssymb,
nofootinbib,eqsecnum,12pt,floatfix]{revtex4-1}

\usepackage{graphicx}
\usepackage{comment}

\newlength{\dummysp}
\newcommand{\tr}{\mathop{{\hbox{Tr} \, }}\nolimits}

\newcommand{\half}{\frac{1}{2}}

\newcommand{\beq}{\begin{eqnarray}}
\newcommand{\eeq}{\end{eqnarray}}

\newcommand{\e}{{\epsilon}}

\newcommand{\vev}[1]{{\langle #1 \rangle}}

\newcommand{\ord}[1]{{{\cal O}(#1)}}
\newcommand{\gappeq}{\mathrel{\rlap {\raise.5ex\hbox{$>$}}
{\lower.5ex\hbox{$\sim$}}}}
\newcommand{\lappeq}{\mathrel{\rlap{\raise.5ex\hbox{$<$}}
{\lower.5ex\hbox{$\sim$}}}}
\newcommand{\myref}[1]{(\ref{#1})}

\newcommand{\ben}{\begin{enumerate}}
\newcommand{\een}{\end{enumerate}}
\newcommand{\bit}{\begin{itemize}}
\newcommand{\eit}{\end{itemize}}

\newcommand{\Ocal}{{\cal O}}

\newcommand{\Zbf}{{\bf Z}}

\def\[{\left [}
\def\]{\right ]}
\def\({\left (}
\def\){\right )}

\begin{document}

\title{Eguchi-Kawai reduction with one flavor of adjoint M\"obius fermion}

\author{William Cunningham}
\email{wjcunningham7@gmail.com}
\affiliation{Department of Physics, Northeastern University, 360 Huntington Ave, 
111 Dana Research Center, Boston MA 02115 USA}

\author{Joel Giedt}
\email{giedtj@rpi.edu}
\affiliation{Department of Physics, Applied Physics and Astronomy,
Rensselaer Polytechnic Institute, 110 8th Street, Troy NY 12065 USA}

\date{December 22, 2015}

\begin{abstract}
We study the single site lattice gauge theory of $SU(N)$ coupled
to one Dirac flavor of fermion in the adjoint representation.  
We utilize M\"obius fermions for this study, and accelerate the calculation
with graphics processing units (GPUs).
Our Monte Carlo simulations indicate that for sufficiently
large inverse 't Hooft coupling $b = 1/g^2 N$, and for $N \leq 10$ the
distribution of traced Polyakov loops has ``fingers'' that
extend from the origin.  However, in the massless case the distribution of
eigenvalues of the untraced Polyakov loop becomes uniform at large $N$,
indicating preservation of center symmetry in the thermodynamic limit.
By contrast, for a large mass and large $b$, the distribution is highly nonuniform
in the same limit, indicating spontaneous center symmetry breaking.
These conclusions are confirmed by comparing to the quenched case,
as well as by examining another observable based on the
average value of the modulus of the traced Polyakov loop.
The result of this investigation is that with massless adjoint
fermions center symmetry is stabilized and the Eguchi-Kawai reduction
should be successful; this is in agreement with most other studies.
\end{abstract}



\maketitle

\section{Introduction}
Eguchi-Kawai reduction \cite{Eguchi:1982nm} is an attractive idea in which a gauge
theory becomes spacetime volume independent
for a large number of colors $N$ in a $U(N)$ or $SU(N)$ gauge theory.
Originally this was shown by demonstrating that the Schwinger-Dyson
equations (loop equations) for suitably defined Wilson loops are equivalent in two lattice Yang-Mills
theories at large $N$:  a model with an infinite number of sites in all four
spacetime dimensions and a model with a single site.
However, after the appearance of \cite{Eguchi:1982nm}
it was rapidly shown that there is a phase transition at which
the center symmetry ($U(1)^d$ for gauge group $U(N)$ or
$Z_N^d$ for $SU(N)$, where $d$ is the number of
spacetime dimensions) is spontaneously broken \cite{Bhanot:1982sh,Kazakov:1982gh},
invalidating the reduction in the continuum limit.  Various
solutions to this problem have been suggested over the years,
but the one that will occupy us here is the one that
by adding fermions
in the adjoint representation the center symmetry may remain
unbroken for all values of the coupling \cite{Kovtun:2007py}.

If this is successful, some important things can be learned about
quantum gauge field theories in large or infinite spacetime volume
by studying a much simpler system.  The large volume theory (e.g., $K^d$
sites in $d$ dimensions)
and the small volume theory (e.g., one site in $d$ dimensions)
have a parent-daughter relationship under an orbifold
by a discrete translation group $\Zbf_K^d$ \cite{Kovtun:2007py}.
The basic observables in the parent theory are single trace
operators which are averaged over all sites of the lattice, e.g.,
\beq
\Ocal_{\text{parent}} =
\sum_x \tr ( U_\mu(x) U_\nu(x + a \hat\mu)
U_\mu^\dagger(x + a \hat\nu) U_\nu^\dagger(x) )
\label{parop}
\eeq
whereas in the daughter theory one has simply 
\beq
\Ocal_{\text{daughter}}
= \tr ( U_\mu U_\nu U_\mu^\dagger U_\nu^\dagger )
\eeq
as the corresponding
operator.  Connected ``correlation functions'' of this class of
operators are equal to each other in the large $N$ limit:
\beq
\lim_{N\to\infty} (K^d)^{M-1} \vev{ \Ocal_1 \cdots \Ocal_M}_{\text{conn.}}^{N,KL}
= \lim_{N\to\infty} \vev{ \Ocal_1 \cdots \Ocal_M}_{\text{conn.}}^{N,L}
\eeq
Here $KL$ is the lattice extent in each direction for the parent (left-hand side),
whereas $L$ is the extent for the daughter (right-hand side).
In the single site lattice case that we consider, $L=a$, the lattice spacing.
For $M=2$ these are susceptibilities in the large volume theory,
and so we could learn about critical exponents by studying a single site lattice.
However, correlations between operators on different timeslices are not
accessible due to the sum over Euclidean time $t$ implied by \myref{parop}, and so the usual
spectral studies could not be performed by this approach.

In this article we examine the proposal of ``center stabilization'' (i.e., the
absence of center symmetry breaking)
by adjoint representation fermion flavors nonperturbatively by Monte Carlo
simulations, restricting ourselves to the case of a single flavor
of Dirac fermion in the adjoint representation (two Majorana flavors).
We will show that for finite $N$ and sufficiently
large inverse 't Hooft coupling $b = 1/g^2 N$ there is the
emergence of ``structure,'' both in the distribution of traced Polyakov loops
and in the eigenvalue distribution of the untraced Polyakov loops.
If the structure is heavily weighted toward nonuniformity,
then center symmetry may be spontaneously broken.  However,
we show that some weak structure still allows for center
symmetry to remain intact, due to tunneling phenomena.  Drawing
the distinction between these two scenarios requires the examination
of a variety of observables.  Another theme of the present paper is that
it is only possible to draw conclusions about spontaneous
breaking of the $Z_N$ center symmetry in the thermodynamic
limit.  In the case of the single-site theory, this corresponds
to $N \to \infty$, since that is the only way to have the
requisite infinite number of degrees of freedom.  We therefore
perform fits to our eigenvalue distribution and extrapolate
to this limit.  Remarkably, we find that for the massless
theory (which is easily obtained from our choice of lattice
fermions by setting the bare mass to zero), the eigenvalue
distribution becomes uniform as $N \to \infty$ for all of the
values of $b$ that we are able to access.  We therefore reach
the conclusion that the center symmetry is certainly stabilized
and the Eguchi-Kawai reduction should be successful in this
theory.  This is consistent with
the continuum one-loop effective potential in the $N \to \infty$ limit,
which indicates
that the eigenvalues of the (untraced) Polyakov loop are uniformly
distributed, and hence center symmetry is unbroken \cite{Kovtun:2007py}.
Also, the single site lattice study of \cite{Bringoltz:2009kb}, which
used Wilson fermions, found that for a certain range of masses
center symmetry was unbroken.  They used
a different discretization, which had the potential to have important effects on
irrelevant operators which change the symmetry at a given choice of $m,b,N$.  
See the related discussion in \cite{Bringoltz:2009mi}.  It is therefore
significant to obtain a similar result using a different lattice fermion.
Furthermore, our analysis of the eigenvalues of the untraced Polyakov
loop is a new ingredient, which we believe puts the fate of
center symmetry on firmer, more quantitative ground.
In \cite{Hietanen:2012ma} the distribution of
\beq
{\tilde P}_\mu = \half \( 1 - \frac{1}{N^2} | \tr U_\mu |^2 \)
\label{plmeas}
\eeq
was studied as an indicator whether or not center symmetry is broken.
It was argued that it should be approximately $1/2$ if the theory is symmetric,
since the Polyakov loop $\tr U_\mu$ will be distributed close to zero.
In Fig.~1 of \cite{Hietanen:2012ma} it can be seen that $P_\mu = 1/2$
to a very good approximation for the $b$ and $N$ values that they
have studied in the one Dirac flavor theory.  It is important to
note that they use the Wilson kernel and a value of the Wilson mass (what we call
$m_5$ below) that has much larger magnitude than the one in our study.
(We also explore the observable ${\tilde P}_\mu$ in the study below.)
So again, various discretizations are leading to a similar conclusion.
The lack of symmetry breaking found in all of these studies is consistent with the findings of the
lattice perturbation theory calculation of \cite{Hietanen:2009ex} for $b=1$
(see Fig.~4 of that paper).
By contrast, the analysis of Ref.~\cite{Lohmayer:2013spa},
which is a sort of semi-classical approach that only keeps the eigenvalues
of the link matrices and throws away the rest of the gauge field information,
indicates that center symmetry is broken at large enough $b$.

\section{Formulation}
The action consists of a gauge part and a fermion part:  $S = S_g + S_f$, where
we use the Wilson plaquette gauge action and a M\"obius Dirac operator.  The latter
is five dimensional, like domain wall fermions, with the four ordinary
spacetime dimensions reduced to a single site.  For the purposes of
the Monte Carlo simulation, we only need the fermion measure, 
and for this the manipulations of \cite{Brower:2012vk} are essential for
reducing this to the determinant of a version of the four-dimensional
overlap operator---one involving the M\"obius kernel rather than
the more conventional Wilson kernel.
Explicitly, we compute the determinant of the overlap operator
\beq
D_{\text{ov}} = \frac{1 + m}{2} + \frac{1 - m}{2} \gamma_5 \e_{L_s}(H_5) 
\label{dov}
\eeq
where for M\"obius fermions the polar approximation to the sign function is used,
\beq
\e_{L_s}(H_5) = \frac{(1+H_5)^{L_s} - (1-H_5)^{L_s}}{(1+H_5)^{L_s} + (1-H_5)^{L_s}}
\label{signf}
\eeq
The Hermitian operator $H_5$ is $\gamma_5$ times the M\"obius kernel,
\beq
H_5 = \gamma_5 \frac{(b_5+c_5) D_W(m_5)}{2 + (b_5-c_5) D_W(m_5)}
\label{h5def}
\eeq
It involves a M\"obius transformation of the Wilson kernel $D_W$, which in turn
depends on the bare ``mass'' $m_5$.
This ``mass'' is also known as the ``domain wall height'' and it does not
correspond to a physical mass, but is rather a parameter of the regulator.
In \myref{dov} the parameter $m$ is the fermion mass in the sense that the partially
conserved axial current mass will
vanish as $m \to 0$, although of course it still requires a
multiplicative renormalization in order to obtain a physical quantity.  
However, the important point is that we can easily obtain the massless
theory by taking $m \to 0$, since there is no additive renormalization (in
the $L_s \to \infty$ limit).
In our studies we have taken
\beq
b_5 = 1.5 , \quad c_5 = 0.5 , \quad m_5 = -1.5
\eeq
based on findings in \cite{Brower:2012vk},
and we have considered various values of $L_s$.  For us increasing $L_s$ does
not increase our cost significantly (since we diagonalize $H_5$ using
dense matrix algorithms), so we take it to be large enough
to have a negligible residual mass.  For $N = 8$ and $L_s = 64$, the residual mass
is $m_{\text{res}} = \ord{10^{-8}}$, which is quite acceptable, so we use
this value in all that follows.

\section{Simulation}
We begin by thermalizing the gauge fields over a period of 
2,500 to 5,000 Monte Carlo iterations, and measurements are subsequently 
taken every 10 iterations, accumulating a total of 10,000 to 100,000 samples,
where the number taken depends on what we found necessary for proper
sampling of the $N$ ground states, corresponding to the $Z_N$ symmetry.
Randomization and rethermalization is performed every 100 samples
in order to overcome the free energy barriers between the $N$ ground states.
Various steps of numerical linear algebra are
required:  we compute the inverse in \myref{h5def} and diagonalize this
operator to compute the sign function in \myref{signf}.  After rotating
back to the original basis, we form the operator \myref{dov}
and compute its determinant by an LU decomposition.
For all of these manipulations we accelerate the calculations
with NVIDIA C2075 graphics processing units (GPUs), making use of the MAGMA library \cite{magma}.  
Our simulation proceeds along the lines of the method used in \cite{Bringoltz:2009kb}.
We update the gauge fields using the Metropolis algorithm, with $S_f = -\ln \det D_{\text{M\"obius}}$
for the fermion effective action, and the Wilson plaquette gauge action.  In doing
this the new fields are chosen by $U_\mu \to V_\mu U_\mu$ with $V_\mu$ a random
$SU(N)$ matrix.  The random matrix is generated using standard $SU(2)$ subgroup
methods of Cabibbo and Marinari, and Okawa \cite{Cabibbo:1982zn,Okawa:1982ic}, though we do not use
the heatbath method that is described in Cabibbo and Marinari---due to the
fermion determinant.
In order to keep acceptance rates to a reasonable level (70\% to 90\% in our runs),
we find that $V_\mu$ must not be too far from a unit matrix.  We use a Gaussian
distribution about the unit matrix, and adjust the variance
as a function of $b$, since the acceptance rate is a sensitive
function of this parameter.  In fact, holding the acceptance rate fixed,
the variance of the Gaussian
distribution is a rapidly decreasing function of $b$, so that
this becomes our chief limitation in going to larger $b$.  If
the variance is too small, we have autocorrelation times that
are absurdly long.  Only one SU(2) subgroup is used per update, 
though all four gauge matrices are updated each iteration.  Autocorrelations 
are monitored by observing how well the Polyakov loops sample the full range
of possible values, as well as trying different block sizes in
our jackknife analysis of errors in the untraced Polyakov loop eigenvalue
distribution.  In particular, in the cases where the distribution of
traced Polyakov loops shows long ``fingers,'' we
check that all $N$ vacua appear in the distribution, which requires
an adequate tunneling between minima of the free energy.  Unitarity of the gauge 
matrices is monitored, due to the possibility of accumulated roundoff error
over our very lengthy runs.
In practice we never found an instance of unitarity being violated beyond
our tolerance of $1.0 \times 10^{-10}$.

\section{Traced Polyakov loop distribution}
\begin{figure}
\begin{center}
\includegraphics[width=4in]{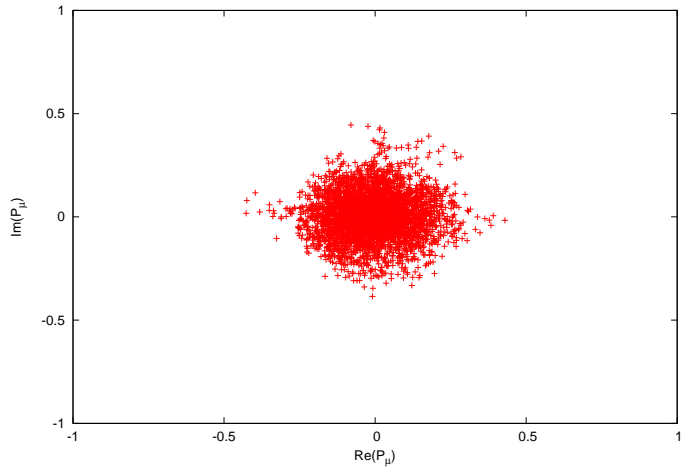}
\caption{A traced Polyakov loop distribution that only has a central lump: $N=6$, $b=0$, $m=0.0001$. \label{symmetric} }
\end{center}
\end{figure}

In this section we explore the behavior of the traditional observable used for the
examination of the fate of the $Z_N$ center symmetry, as a function of $N$ and $b$ 
with $m=0.0001$ which is essentially a massless fermion.
What we use here is the traced Polyakov loop in each direction, which for a single
site lattice is just $P_\mu = \tr U_\mu$.  Traditionally, when the distribution of $P_\mu$ forms
a lump centered at zero, the phase is interpreted as center symmetric.  When the distribution
forms $N$ islands away from zero, the center symmetry is typically interpreted as broken.
In fact we will never find distinct islands, but always find a nonzero density of states in the center
of the diagram.  Rather what we find is that in some cases there is only 
the central lump (Fig.~\ref{symmetric}),
whereas in other cases the lump has ``fingers'' (Fig.~\ref{broken}).  As will be discussed more
at length in Section \ref{spontaneous} below, the nonzero density of states in this central
region allows for a nonzero tunneling rate between the $N$ different ground states.
In this case, the center symmetry may not actually be broken.  
As is also discussed in Section \ref{spontaneous},
we only really expect spontaneous symmetry breaking in the large $N$ limit on the single
site lattice, because the thermodynamic limit must be taken before infinite barriers can
arise between ground states.

We have studied $N=3$ to $10$ and have found the
``critical'' value $b_c$ of $b$ above which fingers form on the distributions in each case.
The results for $m=0.0001$ are shown in Fig.~\ref{grows} and it can be
seen that $b_c$ increases with $N$.  
The simulations become more difficult as $b$ and $N$ increase.  In the case of $N=10$,
and only in this case, this gave rise to slightly ambiguous results for the largest value of $b$.
For instance we found that $b=0.3$, $N=10$ was clearly fingered, whereas the case of $b=0.4$, $N=10$ can be
classified as ``sort-of fingered'' because it suffered from
significant autocorrelation and only appeared to explore two of the ten ground states,
giving rise to two ``fingers'' far away from the origin.

\begin{figure}
\begin{center}
\includegraphics[width=4in]{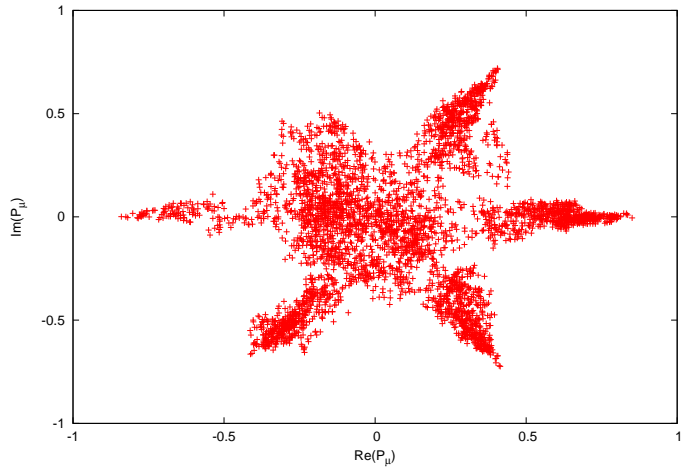}
\caption{A traced Polyakov loop distribution with ``fingers'' $N=6$, $b=0.30$, $m=0.0001$. \label{broken} }
\end{center}
\end{figure}

It is interesting to observe what happens to the distribution of Polyakov loop values
as $b$ is increased.  In Fig.~\ref{N3btrend} it can be seen that for $N=3$ as $b$ is increased
the Polyakov loop distribution moves out into the fingers and away from the center.
Thus it becomes less and less likely that a configuration will tunnel from one of the
fingers into another.  This is an indication that the eigenvalue distribution of the
link variables is far from uniform in the large $b$ limit.

\begin{figure}
\begin{center}
\includegraphics[width=4in]{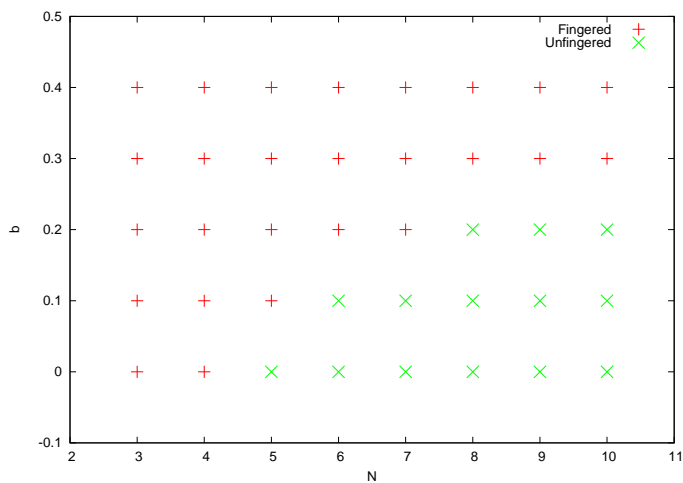}
\caption{The ``critical'' values $b_c$, where fingers clearly form,
increases with the number of colors $N$.  These
results are for $m=0.0001$, essentially a massless fermion.  \label{grows} }
\end{center}
\end{figure}

\begin{figure}
\begin{center}
\begin{tabular}{cc}
\includegraphics[width=3in,height=3in]{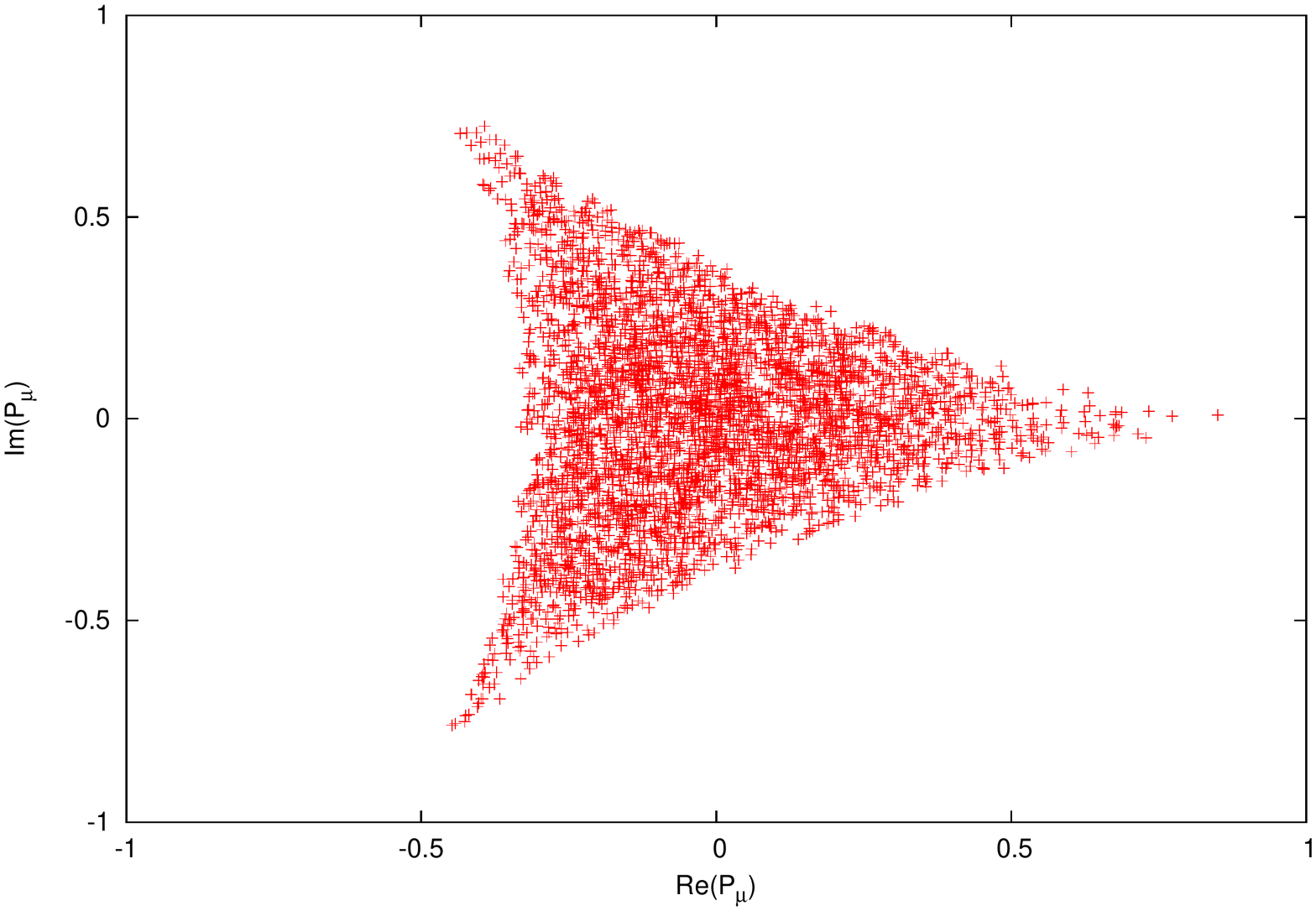}
& \includegraphics[width=3in,height=3in]{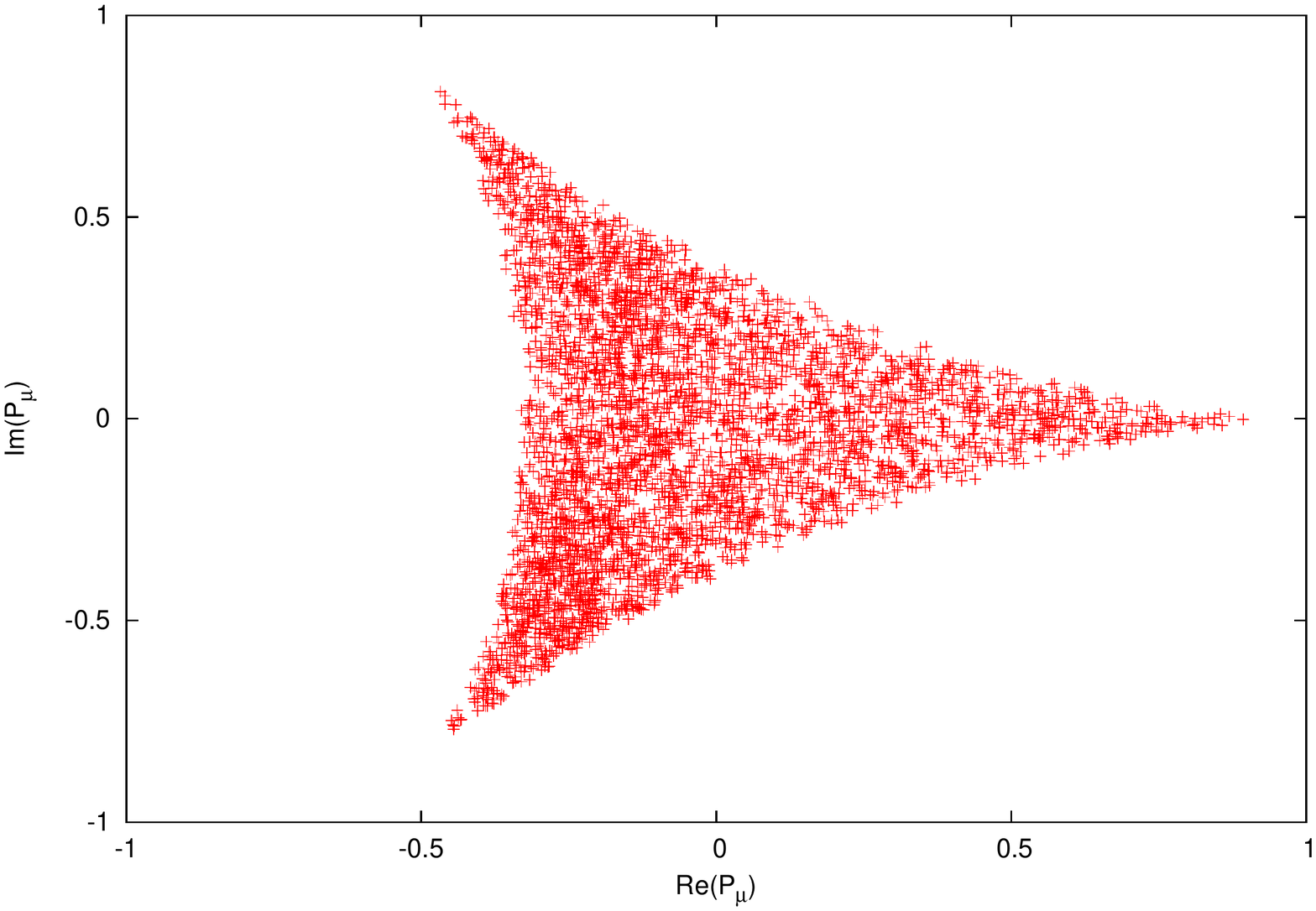} \\
\includegraphics[width=3in,height=3in]{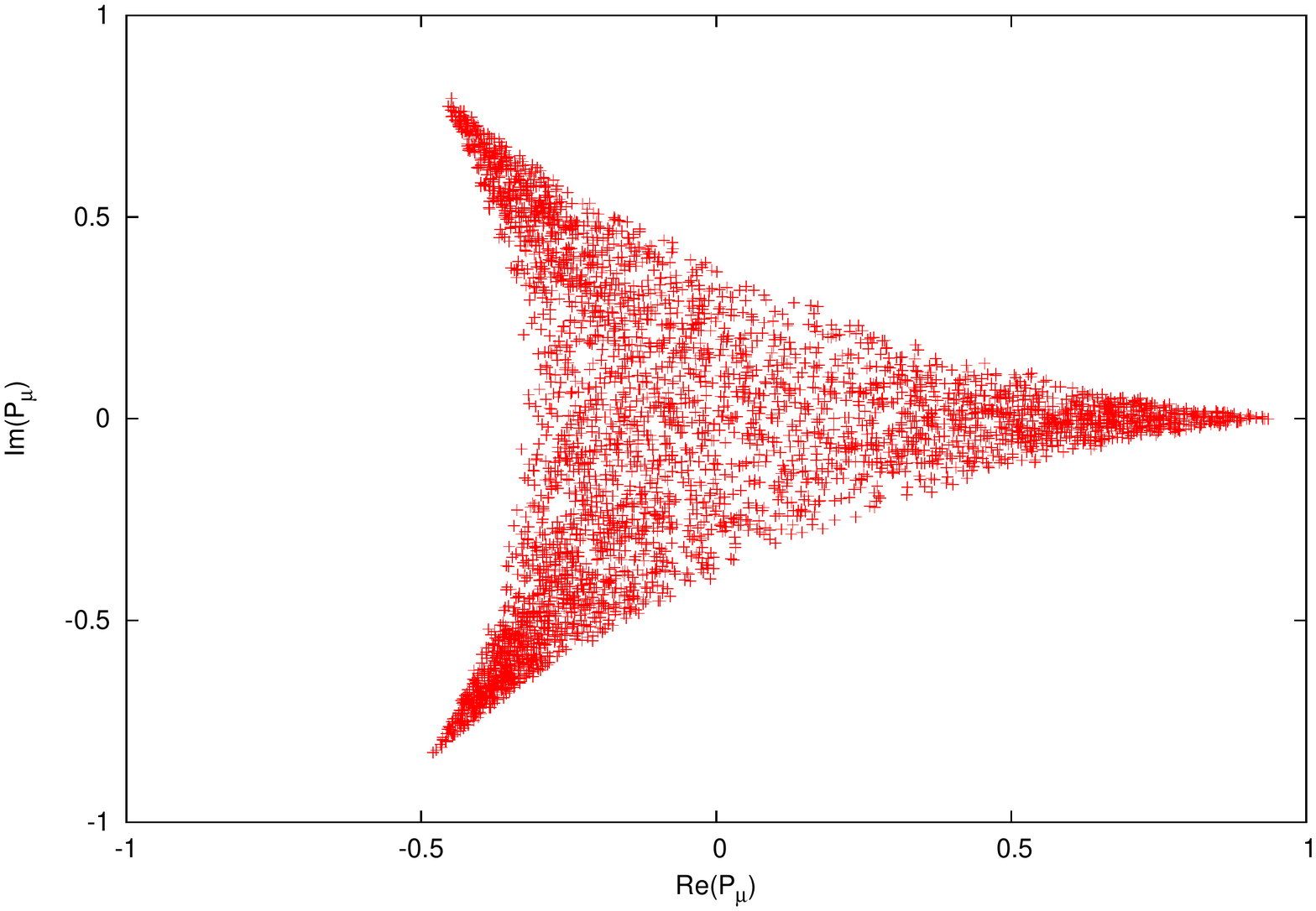}
& \includegraphics[width=3in,height=3in]{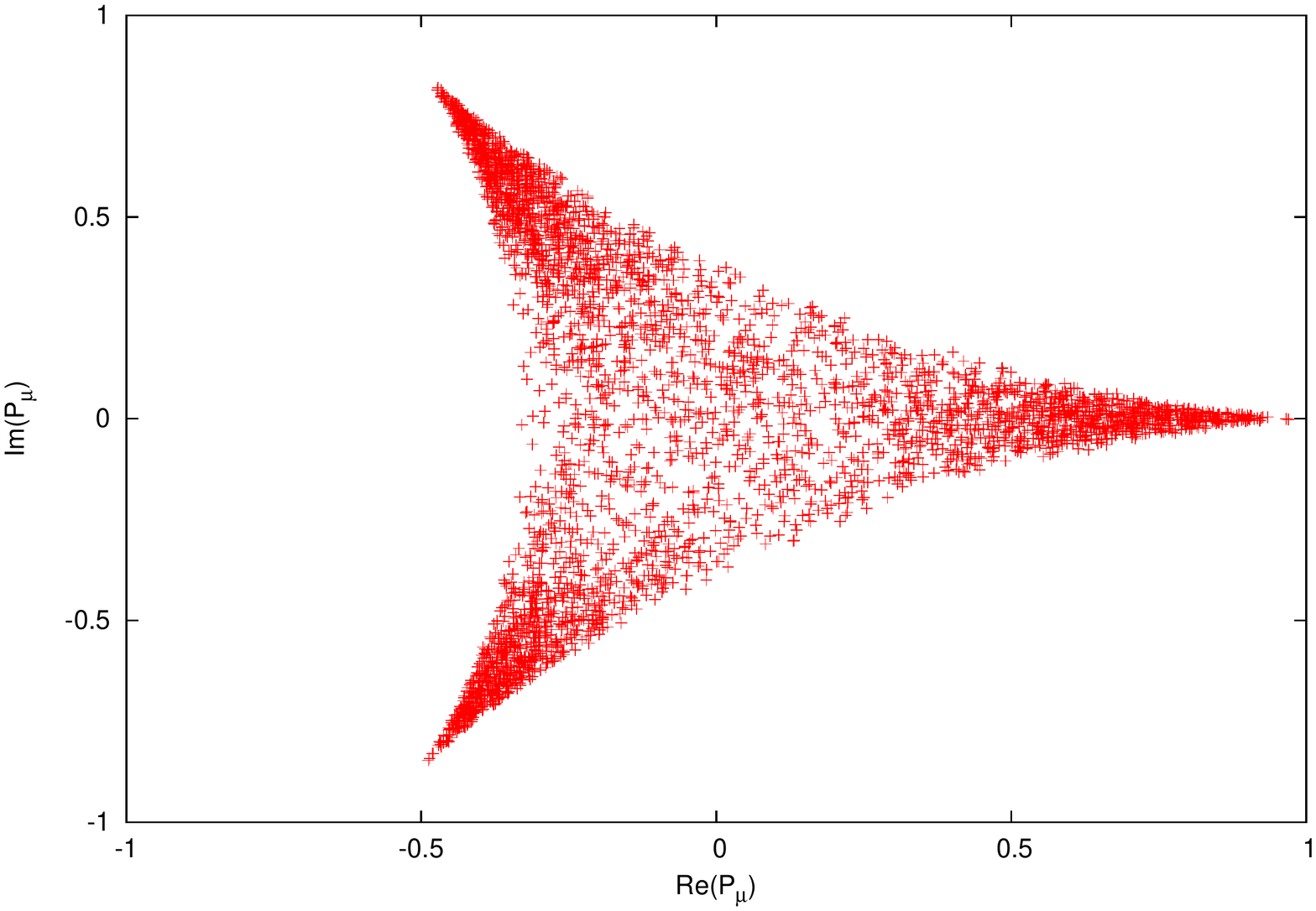}
\end{tabular}
\caption{$N=3$ Polyakov loop distribution for increasing $b$ values.  In the upper left-hand panel
is shown $b=0.00$, in the upper right $b=0.10$, in the lower left $b=0.20$ and in the
lower right $b=0.30$.  It can be seen that the values are tending to move further out
into the three fingers, and away from the center, as $b$ is increased. \label{N3btrend} }
\end{center}
\end{figure}

\section{Eigenvalue analysis}
\label{eva}
If the eigenvalues of the untraced
Polyakov loop operator have a uniform distribution in the $N \to \infty$
limit,\footnote{Since we are studying $SU(N)$, the Haar measure is not
uniform except at $N \to \infty$, but rather has $N$ peaks.  The Haar measure
certainly corresponds to the center symmetric phase since it is
the $b=0$ quenched theory.  Thus simply seeing structure in the distribution
at finite $N$ is not a proof of broken center symmetry.} then the theory
is certainly center symmetric.  These
eigenvalues lie on the unit circle in the complex plane, and are
thus of the form $e^{i\theta}$.

\subsection{Unquenched theory}
\subsubsection{Zero mass}
In our analysis for zero mass ($m=0.0001$ in practice),
the eigenvalue distribution is fit to the following form:
\beq
F(\theta) = A + B \cos(N \theta) + C \cos (2 N \theta)
\label{Fvstheta}
\eeq
In some cases we can set $C \equiv 0$ and still get a good fit; in others,
the $C$ term is necessary.  When $C \equiv 0$, the ratio that we measure
to test uniformity is
\beq
R = \left| \frac{B}{A} \right|
\label{2parR}
\eeq
When $C \not= 0$, the ratio instead is taken to be
\beq
R = \sqrt{ \(\frac{B}{A}\)^2 + \( \frac{C}{A} \)^2 }
\label{3parR}
\eeq
Since we only expect the distribution to become uniform in the $N \to \infty$
limit, the important thing is how $R$ depends on $N$.  Thus we fit the ratio to
\beq
R(N) = c_0 + \frac{c_{-1}}{N} + \frac{c_{-2}}{N^2} + \cdots
\label{RvsN}
\eeq
and find good agreement for each value of $b$; of course the coefficients
depend on $b$.

A comment here is in order.  In \myref{RvsN} we fit the data to a smooth
function of $N$.  However in the phase diagram Fig.~\ref{grows} we make
a binary distinction between fingered and unfingered.  In fact as one
moves toward increasing $N$ at fixed $b$, the fingers gradually shrink
and eventually one ends up with a central blob.  Thus the transition is
in this sense smooth, and the classification into fingered and unfingered
does not reflect a discontinuity.  Indeed, it is a crossover behavior,
and the ``critical'' $b_c$ does not indicate a singularity of any kind.

\begin{figure}
\begin{center}
\includegraphics[width=4in]{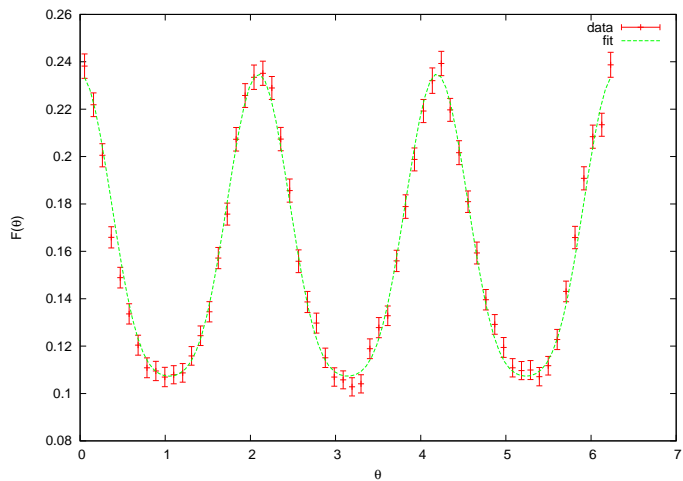}
\caption{Eigenvalue distribution for $N=3$, $b=0.10$, $m=0.0001$, and three
parameter fit \myref{Fvstheta}.  The $\chi^2/\text{d.o.f.}$ was $1.04$.
Errors in the eigenvalue distribution
were estimated with jackknife elimination of blocks of size 500.
\label{N3fit} }
\end{center}
\end{figure}

\begin{figure}
\begin{center}
\includegraphics[width=4in]{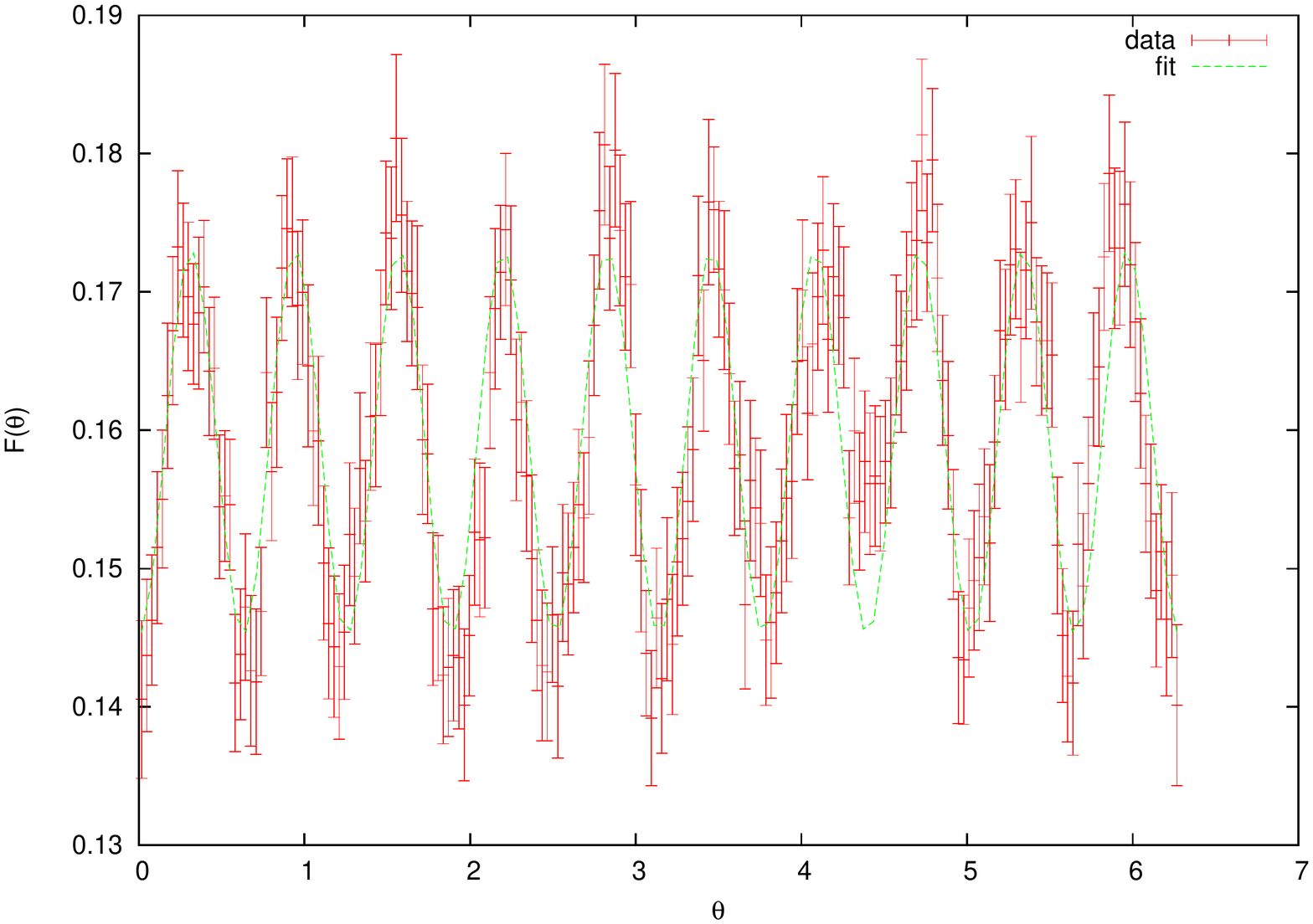}
\caption{Eigenvalue distribution for $N=10$, $b=0.10$, $m=0.0001$, and three
parameter fit \myref{Fvstheta} with $C \equiv 0$.  The $\chi^2/\text{d.o.f.}$ was $0.96$.
Errors in the eigenvalue distribution
were estimated with jackknife elimination of blocks of size 500.
\label{N10fit} }
\end{center}
\end{figure}

\begin{figure}
\begin{center}
\includegraphics[width=4in]{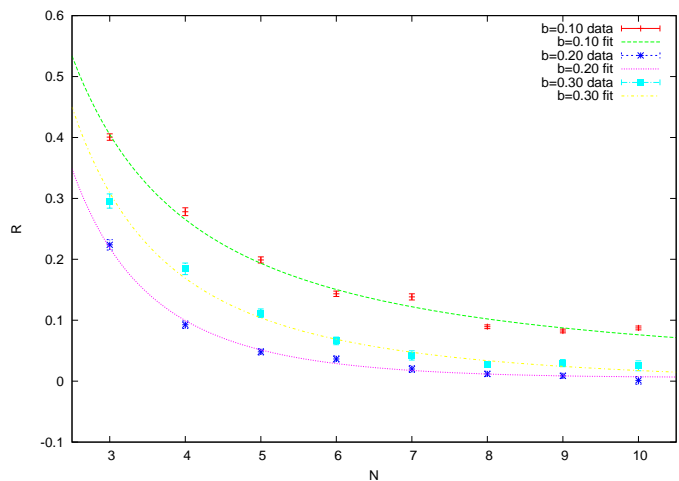}
\caption{The ratios \myref{2parR} or \myref{3parR}
versus $N$, and the fits, for $b=0.10, 0.20$ and $0.30$.
Fit results are summarized in Table \ref{tabm0001ratfit}.
All three curves appear to tend toward zero in the large
$N$ limit, corresponding to a uniform distribution.
This is indicative of unbroken center symmetry in
the thermodynamic limit.  \label{m0001ratfit} }
\end{center}
\end{figure}

For $b=0.10$, only $N=3$ required the three parameter fit \myref{Fvstheta}.  For all other
values of $N$ we were able to set $C=0$ and obtain a good fit.  
Examples are shown in Figs.~\ref{N3fit} and \ref{N10fit}.
The subsequent fit to \myref{RvsN}
is shown in Fig.~\ref{m0001ratfit}, with the fit parameters obtained 
displayed in Table~\ref{tabm0001ratfit}.
The value of the constant term $c_0$ is consistent with zero, indicating that the
eigenvalue distribution becomes uniform in the large $N$ limit.  Thus we find
that the center symmetry is certainly unbroken in the thermodynamic limit in the case of $b=0.10$.

\begin{table}
\begin{center}
\begin{tabular}{ccccc}
$b$ & $c_0$ & $c_{-1}$ & $c_{-2}$ & $\chi^2/\text{d.o.f.}$ \\ \hline
0.10 & -0.006(38) & 0.65(43) & 1.8(1.0) & 3.16 \\
0.20 & 0.033(17) & -0.60(19) & 3.48(46) & 1.08 \\ 
0.30 & -0.011(5) & --- & 2.88(14) & 1.64 \\ \hline
\end{tabular}
\caption{Results of fits of the ratios \myref{2parR} and \myref{3parR}
to \myref{RvsN}.  In the case of $b=0.30$ we are able to
set $c_{-1}=0$.  All three fits give results that
are $\sim 2 \sigma$ consistent with a uniform
distribution in the large $N$ limit.   \label{tabm0001ratfit}}
\end{center}
\end{table}

For $b=0.20$, data for $N=3,\dots,7$ required all three parameters in \myref{Fvstheta},
whereas for $N=8,9,10$ we were able to set $C \equiv 0$, since the two parameter
fit was acceptable and $C$ was very small if it was included.  The fit to
\myref{RvsN} is shown in Fig.~\ref{m0001ratfit}, and the fit parameters 
are tabulated in the second row of Table \ref{tabm0001ratfit}.
The value of the constant term, $c_0$, is $1.9\sigma$ from zero, which we
view as most likely consistent with a uniform distribution in the large $N$ limit, given
the uncertainties in the measurement [e.g., a fairly simple-minded functional
form has been assumed in \myref{Fvstheta}.]  Thus we conclude that for $b=0.20$,
the center symmetry is probably unbroken in the thermodynamic limit.

For $b=0.30$, the fits to the eigenvalue distribution required three parameters:
The ratio \myref{3parR} was then fit to \myref{RvsN}
with $c_{-1} \equiv 0$ since it was found that a $1/N$ term did not improve the fit.  The result
was row three of Table \ref{tabm0001ratfit}, and
shown in Fig.~\ref{m0001ratfit}.
Since a negative value in the $N \to \infty$ limit is actually excluded by the
positive definite form of $R$, it is clear that the value of $c_0$ is merely
a fitting error; the interpretation is that it is actually zero, since
it is close to zero ($2.2\sigma$), consistent with a uniform distribution
in the $N \to \infty$ limit.  We therefore conclude that for $b=0.30$,
the center symmetry is most likely unbroken in the thermodynamic limit.

We were unable to obtain reliable results for the eigenvalue distribution
for larger values of $b$ because
the acceptance rates in the simulation are tending to zero, forcing very
small moves, which leads to enormous autocorrelation times and incomplete
sampling.  The eigenvalue distributions require many more statistically
independent samples in order to get good fits than does the analysis
of whether or not the traced Polyakov loop has fingers, which
is why we were only able to go to $b=0.30$ in the former case
but were able to go to $b=0.40$ in the latter case.  This is
also true of the ${\tilde P}_\mu$ observable considered below
in Sec.~\ref{sec:ptilde}.
All the values of $b$ that we have been able to examine
in this massless case indicate a center symmetric phase in the large $N$
limit.  We thus conclude that there is no evidence of center symmetry
breaking from the perspective of the eigenvalue distribution in the
massless theory.

\subsubsection{Nonzero mass}
\begin{figure}
\begin{center}
\includegraphics[width=4in]{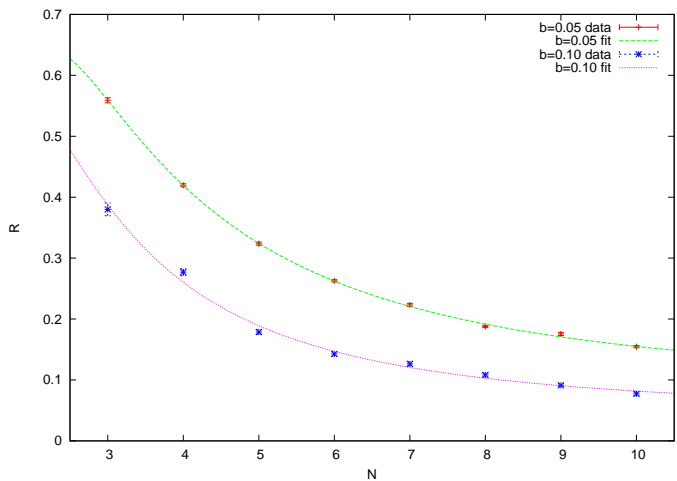}
\caption{The ratios versus $N$, and the fits, for $b=0.05, 0.10$ and $m=0.1$.  \label{b05b10m10fit} }
\end{center}
\end{figure}

In order to provide some contrast, we next consider the case of nonzero mass. 
If the mass is large enough, we should recover the quenched result of
center symmetry breaking starting around $b \sim 0.15$. 
For $b=0.05$ and $m=0.1$, we find the two parameter fit to the distribution
is successful, and that the ratio must be fit to
\beq
R(N) = c_0 + c_{-2} N^{-2} + c_{-3} N^{-3}
\label{ratfit1}
\eeq
in order to get good agreement with the data.  The result is
\beq
c_0 = 0.081(3), \quad c_{-2} = 8.8(3), \quad c_{-3} = -13.4(8), \quad \chi^2/\text{d.o.f.} = 1.69
\eeq
and the data and fit are displayed together in Fig.~\ref{b05b10m10fit}.
The constant term (corresponding to the $N \to \infty$ limit) is significantly nonzero
when taking into account the error of the fit.  A nonuniform distribution in the
large $N$ limit is clearly indicated.

We have also looked at larger
$b$ to see if this signal of a nonuniform distribution strengthens.
For this purpose we studied $b=0.10$.  In that case our best fit occurs
with
\beq
R(N) = c_0 + c_{-2} N^{-2} + c_{-4} N^{-4}
\label{ratfit2}
\eeq
with
\beq
c_0 = 0.043(8), \quad c_{-2} = 4.0(5), \quad c_{-4} = -8(5), \quad \chi^2/\text{d.o.f.} = 4.9
\eeq
The data and fit are shown in Fig.~\ref{b05b10m10fit}.
While the $\chi^2/\text{d.o.f.}$ is not very good, we find the nonzero value
of $c_0$ to be robust with respect to other choices of fit.  The data
seem to indicate that also for this value of $b$, there is a nonuniform
distribution in the limit of infinite $N$, though the conclusion is not
any stronger.

\begin{figure}
\begin{center}
\includegraphics[width=4in]{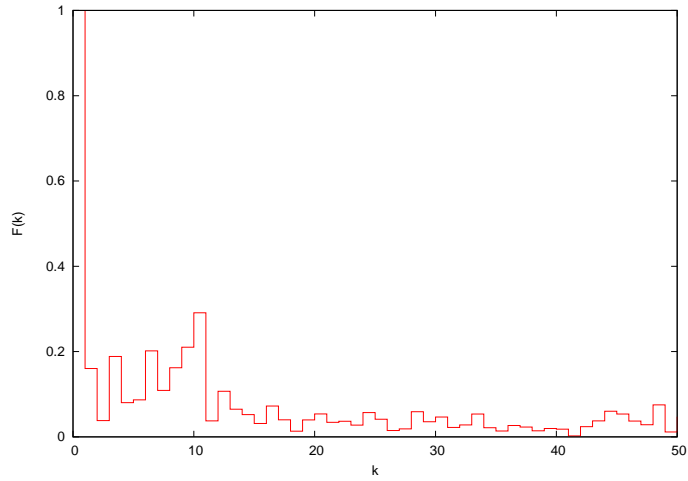}
\caption{Example of the distribution of $F(k)\equiv|{\tilde f_k}|$ for $N=5$, $b=0.20$, $m=0.1$.
Note that $F(0)=15.9$ extends out of the field of view, in order to show the other, much
smaller values.
\label{ftrexample} }
\end{center}
\end{figure}

\begin{figure}
\begin{center}
\includegraphics[width=4in]{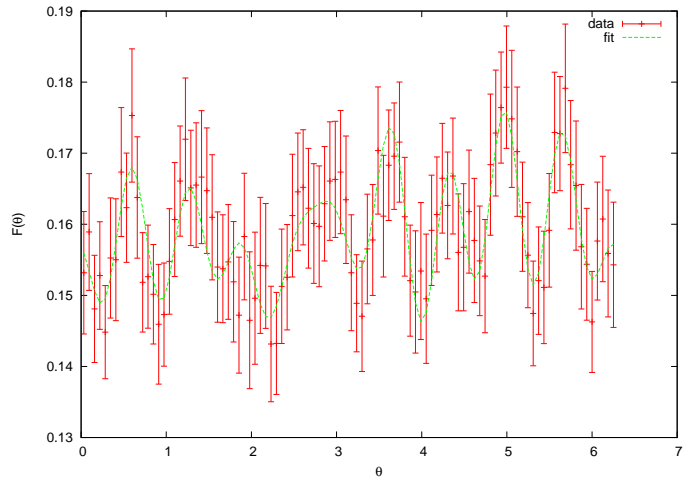}
\caption{Example of the fit to the eigenvalue distribution using Eq.~\myref{genform} for $N=5$, $b=0.20$, $m=0.1$.
Here, $n=6$, the six leading nonzero modes shown in Fig.~\ref{ftrexample}, in addition to the constant
$k=0$ mode.
\label{ftrfitexample} }
\end{center}
\end{figure}

\begin{figure}
\begin{center}
\includegraphics[width=4in,height=3in]{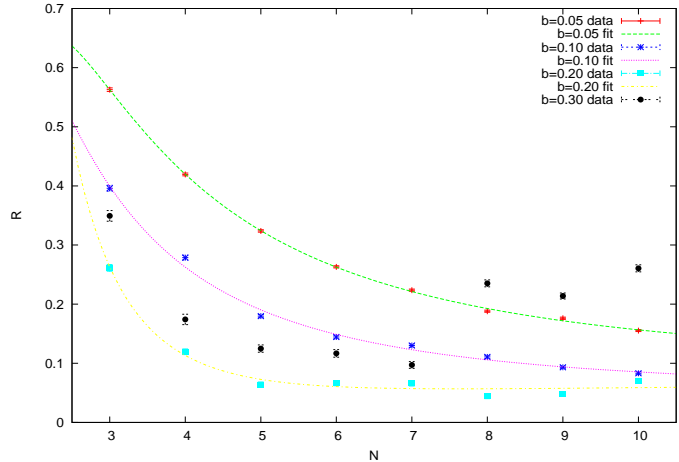}
\caption{The ratio \myref{ratftrn} versus $N$, and the fit, for $b=0.05, 0.10, 0.20, 0.30$, $m=0.1$,
using the Fourier transform method.  Here $n=5$ for $b=0.05, 0.10$ and $n=6$ for
$b=0.20, 0.30$.  \label{m10ftrfit} }
\end{center}
\end{figure}

Moving to $b=0.20$, fits based on \myref{Fvstheta} no longer work well, missing
other modes that are apparent in the data.
The eigenvalue distribution requires a more sophistocated
fitting procedure which we now explain.  We have used $20 N$ bins, with boundaries
at
\beq
\theta_j = j \frac{\pi}{10 N}
\eeq
so that correspondingly we have distribution heights $f_j=f(\theta_j)$.  Next we
perform a discrete Fourier transform of this function to obtain ${\tilde f}_k$,
taking into account that $f_j$ is real (we use FFTW \cite{fftw} for this).
Naturally we find the amplitude $|{\tilde f}_0|$ to be by far the largest, corresponding
to the constant mode.  Next we sort the amplitudes into descending order,
\beq
|{\tilde f}_{k_0}| > |{\tilde f}_{k_1}| > \cdots > |{\tilde f}_{k_n}| > \cdots
\eeq
An example of the amplitudes versus $k$ is shown in Fig.~\ref{ftrexample}.
We then fit the distribution to
\beq
F(\theta) = a_0 + a_1 \cos(k_1 \theta + b_1) + a_2 \cos(k_2 \theta + b_2) + \cdots + a_n \cos(k_n \theta + b_n)
\label{genform}
\eeq
An example is displayed in Fig.~\ref{ftrfitexample}.
For $b=0.05$ and $0.10$ we find that $n=5$ is sufficient, and in fact $n=6$ fails for $b=0.10$ in
that one of the coefficients has over 100\% error.  For $b=0.20$ and $0.30$ we find that the
fits are improved by taking $n=6$.  In either case the ratio is computed from
\beq
R = \[ \(\frac{a_1}{a_0}\)^2 + \cdots + \(\frac{a_n}{a_0}\)^2 \]
\label{ratftrn}
\eeq
which is an obvious generalization of \myref{3parR}.
The results are shown in Fig.~\ref{m10ftrfit}.  Fit results
are given in Table \ref{tabratftr}.

\begin{table}
\begin{center}
\begin{tabular}{cccccc} \hline
$b$ & $c_{p_1}$ & $c_{p_2}$ & $c_{p_3}$ & $\chi^2/\text{d.o.f.}$ & $(p_1,p_2,p_3)$ \\ \hline
0.05 & 0.0835(35) & 8.58(29) & -12.81(84) & 2.58 & (0,2,3) \\ 
0.10 & 0.0482(79) & 3.78(44) & -5.6(3.9) & 8.23 & (0,2,4) \\
0.20 & 0.071(13) & -2.48(99) & 12.6(2.9) & 7.27 & (0,2,3) \\
\hline
\end{tabular}
\caption{Ratio fit results when the Fourier transform method and $n=5$ or $6$ is used.
We use either \myref{ratfit1} or \myref{ratfit2} for the form of $R(N)$,
denoted by the powers $(p_1,p_2,p_3)$ in the last column.  No fit is
performed for $b=0.30$, because as can be seen from Fig.~\ref{m10ftrfit} the
$N=8, 9, 10$ results break away from the trend at smaller $N$.  \label{tabratftr}}
\end{center}
\end{table}

For the cases of $b=0.05, 0.10, 0.20$ we find that the fit indicates
that the eigenvalue distribution is not uniform in the large $N$ limit, in
the first two cases consistent with findings described in previous paragraphs
that did not use the Fourier transform method.
In the case of $b=0.30$, we cannot fit to either of the forms \myref{ratfit1} or \myref{ratfit2}
because there is a jump in the behavior at $N=8$.  This significant nonuniformity
at large $N$ is consistent with what is seen in the quenched
case considered below for $b > 0.15$.  Thus it appears that the quenched behavior
of center symmetry breaking at large $N$ is obtained for
this large mass of $m=0.1$ when $b \gappeq 0.20$.  It is reasonable to assume that the $b$ value
for which this begins to occur is somewhat larger than in the quenched case for a finite mass $m$, as
compared to $m \to \infty$, where the quenched transition of $b \sim 0.15$ would occur.

\subsection{Quenched theory}
\begin{figure}
\begin{center}
\includegraphics[width=4in]{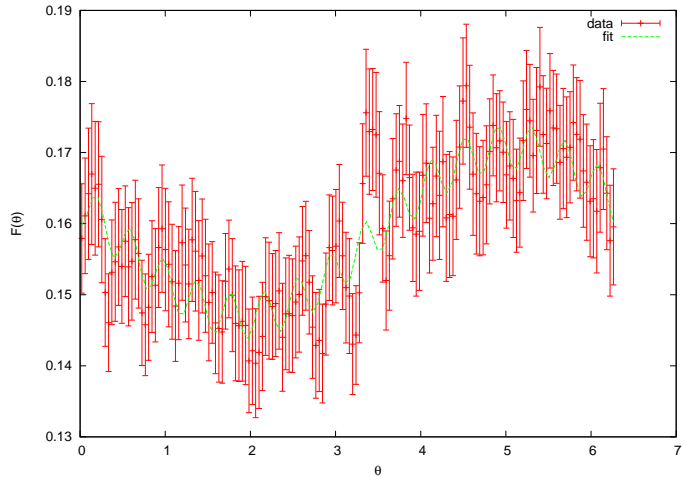}
\caption{Quenched theory eigenvalue distribution for $b=0.3$, $N=8$,
fit to \myref{Fvstheta2}.  The goodness of
the fit was $\chi^2/\text{d.o.f.}=0.40$. Comparing to Fig.~\ref{N10fit},
it can be seen that for the quenched theory at large $N$ with $b$ greater
than the critical value there is a drastic change in the eigenvalue
distribution.  \label{quev1}}
\end{center}
\end{figure}

\begin{figure}
\begin{center}
\includegraphics[width=4in]{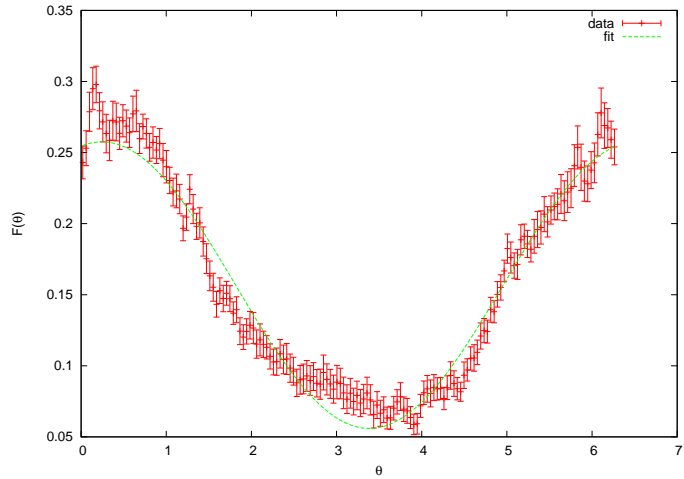}
\caption{Quenched theory eigenvalue distribution for $b=0.4$, $N=8$, 
fit to \myref{Fvstheta2} with $B=0$.  The goodness of
the fit was $\chi^2/\text{d.o.f.}=2.17$. \label{quev2}}
\end{center}
\end{figure}

For purposes of further comparison, we have also analyzed the eigenvalue
distribution in the quenched theory (i.e., setting the fermion
determinant to unity).  The eigenvalue distribution was found
to have a few different forms, in contrast to the massless unquenched case
where \myref{Fvstheta} always worked.  The other forms that were
required were
\beq
F(\theta) = A + B \cos(2 N \theta) + C \cos(\theta + D)
\label{Fvstheta2}
\eeq
and \myref{Fvstheta} with $B = 0$.  In some cases we could also set $B=0$
or $D=0$ in \myref{Fvstheta2} and obtain a good fit (the parameter dropped
was consistent with zero if it was included).  Two examples of
these different shapes are shown in Figs.~\ref{quev1} and \ref{quev2}.

We believe that it is significant that the $N$-fold periodicity of \myref{Fvstheta}
is lost and replaced with \myref{Fvstheta2}, $C \not= 0$, for those values
of $b$ that seem to have broken center symmetry in the large $N$ limit according to the detailed
analysis below.  For $b=0.2$ the alternative form \myref{Fvstheta2} must
be used for $N \geq 6$, for $b=0.3$ it must be used for $N=4$ and $N \geq 6$,
and for $0.4$ it must be used for $N \geq 4$.  Thus the breakdown in $N$-fold periodicity
occurs at smaller and smaller $N$ as $b$ increases, corresponding to a more
dramatic violation of center symmetry.  This also correlates with the
fact that in the massive unquenched case of $b \geq 0.20$ it was necessary
to use more general forms \myref{genform}.

\begin{figure}
\begin{center}
\includegraphics[width=4in]{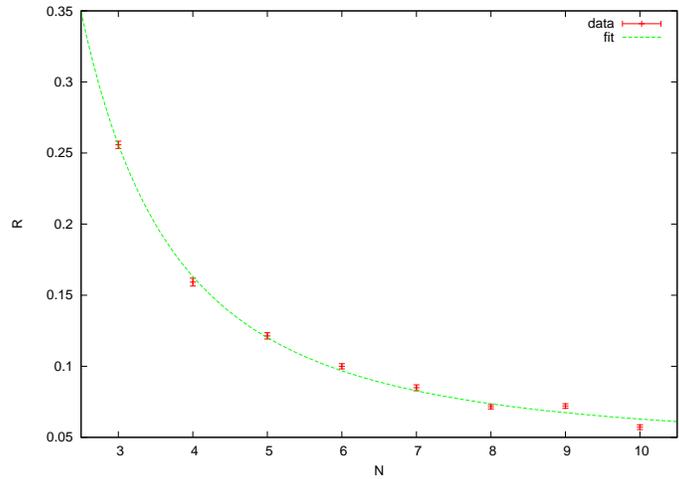}
\caption{Quenched $b=0.10$ ratio versus $N$.  There is a clear
decrease with increasing $N$.  \label{ratio_qu_b10}}
\end{center}
\end{figure}

\begin{figure}
\begin{center}
\includegraphics[width=4in]{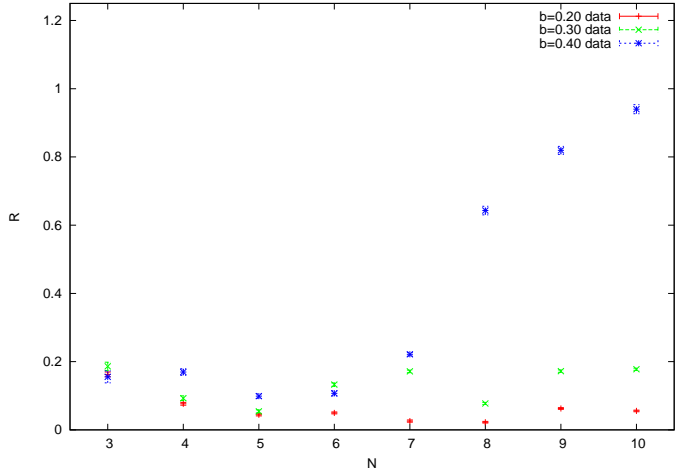}
\caption{Quenched $b=0.20, 0.30, 0.40$ ratio versus $N$.  The ratio either
approaches a constant value significantly different from zero, or shows
an increasing trend with $N$.
Clearly there is a dramatic change for $b \geq 0.40$. \label{ratio_qu_b203040}}
\end{center}
\end{figure}

After the fits were performed, ratios were then formed
using either \myref{2parR} (with $B \to C$ in the case where $B=0$) or 
\myref{3parR} (ignoring the parameter $D$ when \myref{Fvstheta2} was used).
The results are shown in Figs.~\ref{ratio_qu_b10} and \ref{ratio_qu_b203040}. 
Whereas for $b=0.10$ there is a clear decrease with increasing $N$,
for the larger values of $b$ the trend at large $N$ is either to a
constant nonuniformity or one that is rapidly increasing.  Naturally
this strengthens as $b$ is pushed to larger values.

Fitting the ratio versus $N$ for $b=0.1$, using the form 
\beq
R(N) = c_0 + c_{-2} N^{-2}
\label{Rsimp}
\eeq
gives 
\beq
c_0 = 0.0438(22), \quad c_{-2} = 1.909(57), \quad \chi^2/\text{d.o.f.} = 4.88
\eeq
We attempted other more general forms of $N$ dependence; none of these reduced the $\chi^2/\text{d.o.f.}$.
However, there is a clear decrease with $N$ to a small constant in the
large $N$ limit.  Thus the eigenvalue distribution becomes to
a very good approximation uniform in the large $N$ limit.
It will be argued in Section \ref{spontaneous} below that
this is an indication of unbroken center symmetry.  This is also consistent
with old results such as \cite{Bhanot:1982sh} that place $b=0.10$ below the transition.

By contrast, for $b=0.20, 0.30, 0.40$ we cannot fit any smooth function to
the data, so it is not possible to extrapolate to large $N$.  The most
that can be said is that the ratio tends to a small value for $b=0.20$ and
a relatively large value $\sim 0.2$ for $b=0.30$.  For $b=0.40$ the ratio is
monotonically increasing at large $N$, reaching quite large values $\sim 1.0$. 
For purposes of comparison, Ref.~\cite{Bhanot:1982sh}
examined $N=5$ and found fluctuations in the free energy between quadratic and quartic
behavior for what is $b \geq 0.15$ in our language.  
They interpreted this as evidence for symmetry breaking.
This suggests that the irregular
behavior we see in Fig.~\ref{ratio_qu_b203040} is a precursor to
spontaneous symmetry breaking in the large $N$ limit.

The conclusion we draw is that in the quenched theory $b=0.10$ is certainly
center symmetric and $b=0.40$ is certainly broken in the thermodynamic
limit of a large number of colors.  An irregular behavior and significant large $N$ nonuniformity
occurs for $b=0.2$ and $b=0.3$, which is consistent with symmetry
breaking in the large $N$ limit.  Looking at the size of the fluctuations
in the eigenvalue spectrum relative to the constant part gives
a powerful way to distinguish between, in particular, the two cases of $b=0.10$ and $b=0.40$.  Intermediate
values of $b$ are harder to differentiate.

\section{${\tilde P}_\mu$ observable}
\label{sec:ptilde}
In this section we consider the observable in Eq.~\myref{plmeas} above,
which has been used in previous study \cite{Hietanen:2012ma}.  We will
find that it is not decisive in identifying center symmetry breaking,
because it is difficult to make a binary distinction based on the
value of this quantity.  Indeed we will find that the ratio
obtained in the previous section is a much more reliable indicator,
and that the values of ${\tilde P}_\mu$ are merely supportive of
the conclusions reached by that approach, in a suggestive way.

\subsection{Unquenched theory}
\begin{figure}
\begin{center}
\includegraphics[width=4in]{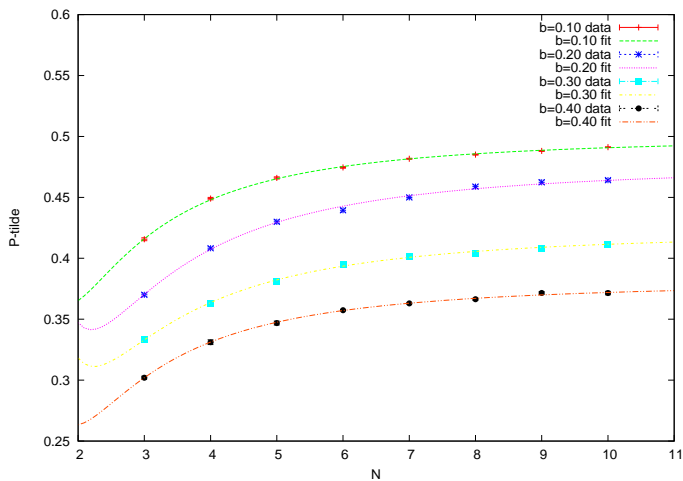}
\caption{${\tilde P}_\mu$ versus the number of colors $N$ for various
values of the inverse 't Hooft coupling $b$, with $m=0.0001$.  It can be seen that as
$b$ is increased, the distribution of Polyakov loops becomes less
central in the large $N$ limit, pushing ${\tilde P}_\mu$ away
from $1/2$. \label{ptilde_m0001} }
\end{center}
\end{figure}

One advantage of the ${\tilde P}_\mu$ observable is that it is
obtained with high accuracy from the simulations, as can be
seen in Fig.~\ref{ptilde_m0001}.  Indeed, the error bars (estimated
with jackknife block elimination as before) are barely 
visible.\footnote{The size of relative errors could be compared to the much larger error bars for instance
in Fig.~\ref{N10fit}; of course, when we form the ratios above
they have much smaller relative error because the large number
of bins in the eigenvalue distribution combine to give small
uncertainty in the fit results.}
In this figure we show ${\tilde P}_\mu$
for four values of $b$ (in this case $b=0.4$ was possible
because the observable has far smaller fluctuation than
the eigenvalue distributions of the previous section) in
the approximately massless case of $m=0.0001$.  We have fit
the data to
\beq
A + B N^{-2} + C N^{-4}
\label{ffque}
\eeq
and in all cases obtain reasonably good agreement, as can
be seen in Table \ref{ptfit_m0001}.

\begin{table}
\begin{center}
\begin{tabular}{ccccc}
\hline
b & A & B & C & $\chi^2/\text{d.o.f.}$ \\
\hline
0.10 & 0.49983(63) & -0.927(44) & 1.55(41) & 1.91 \\
0.20 & 0.4767(17) & -1.308(83) & 3.16(69) & 1.55 \\
0.30 & 0.4223(11) & -1.113(61) & 2.79(49) & 0.88 \\
0.40 & 0.38080(98) & -0.901(53) & 1.72(42) & 0.71 \\
\hline
\end{tabular}
\caption{Fit results for ${\tilde P}_\mu$ in the unquenched theory with $m=0.0001$,
comparing data to Eq.~\myref{ffque}. \label{ptfit_m0001}}
\end{center}
\end{table}

The large $N$ limiting behavior of ${\tilde P}_\mu$ is given by the coefficient $A$.
It can be seen that this quantity declines as $b$ is increased, indicating
that the data for the Polyakov loop is spreading out away from the origin
in the complex plane and is becoming less central.  However, it is impossible
to tell from ${\tilde P}_\mu$ whether or not barriers are emerging between
the $N$ vacua, so one cannot draw a firm conclusion about spontaneous center symmetry breaking.  In fact, since
the eigenvalue analysis above indicated that center symmetry is not broken
in the unquenched case, a consistent interpretation would require that the
decrease in ${\tilde P}_\mu$ is not related to barrier formation, but only
a less centralized distribution in the Polyakov loop which still allows
tunneling in the large $N$ limit, destroying any putative order.

\begin{figure}
\begin{center}
\includegraphics[width=4in]{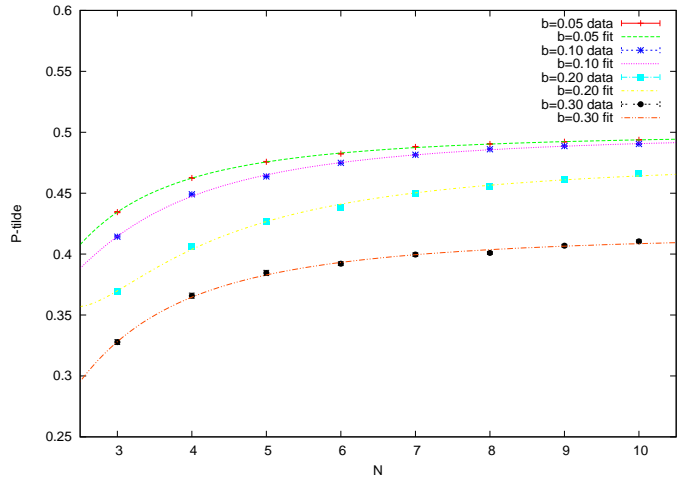}
\caption{${\tilde P}_\mu$ versus the number of colors $N$ for 
the inverse 't Hooft coupling $b=0.05, 0.10, 0.20, 0.30$ that we
ran with $m=0.1$.   \label{ptilde_m1000} }
\end{center}
\end{figure}

We have also computed ${\tilde P}_\mu$ for the massive case $m=0.1$ that
was considered previously, with results shown in Fig.~\ref{ptilde_m1000}.
Comparing the $b=0.10$ entry of Table~\ref{ptfit_m1000} to the $b=0.10$
entry of Table~\ref{ptfit_m0001} it can be seen that the mass has
little effect on the ${\tilde P}_\mu$ observable in the large $N$ limit for small values of $b$.  Both $b=0.05$
and $b=0.10$ with the large mass $m=0.1$ 
extrapolate to $1/2$ to a very good approximation in the large $N$ limit,
indicating an absence of spontaneous symmetry breaking.  We also
see that the large $N$ limit, given by the coefficient $A$, is
quite similar between the massless and massive cases for $b=0.20, 0.30$.  The eigenvalue analysis indicated
that $b=0.30$ was most likely broken, and there was a clear distinction
between that behavor for $m=0.0001$ versus $m=0.1$.  By contrast
the ${\tilde P}_\mu$ observable does not really allow for a way
to differentiate between the massless versus large mass scenario in
this regime where we expect that the center symmetry is spontaneously
broken in the latter case.

\begin{table}
\begin{center}
\begin{tabular}{ccccc}
\hline
b & A & B & C & $\chi^2/\text{d.o.f.}$ \\
\hline
0.05 & 0.49988(24) & -0.616(17) & 0.25(16) & 1.53 \\
0.10 & 0.49978(55) & -0.928(38) & 1.45(36) & 2.40 \\
0.20 & 0.4784(16) & -1.470(91) & 4.44(77) & 1.70 \\
0.30 & 0.4175(18) & -0.90(11) & 0.85(92) & 1.74 \\
\hline
\end{tabular}
\caption{Fit results for the unquenched theory with $m=0.1$,
comparing data to Eq.~\myref{ffque}. \label{ptfit_m1000}}
\end{center}
\end{table}

\subsection{Quenched theory}
\begin{figure}
\begin{center}
\includegraphics[width=4in]{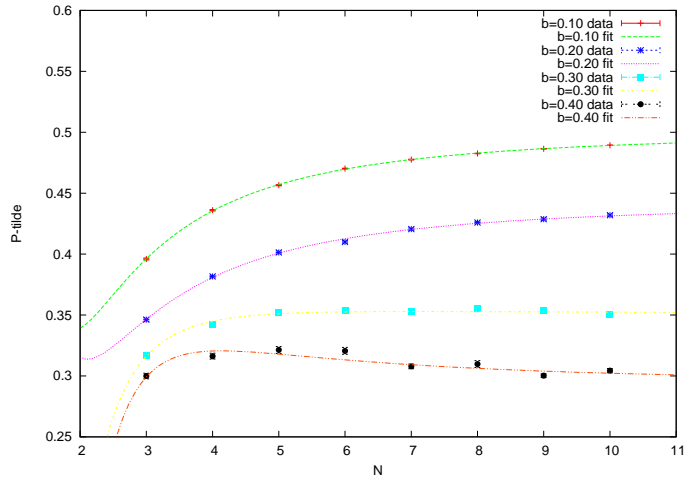}
\caption{${\tilde P}_\mu$ versus the number of colors $N$ for various
values of the inverse t' Hooft coupling $b$ in the quenched case.  The values for large
$N$ (i.e., the value of $A$) are consistent with center
symmetry breaking in the cases of $b=0.30$ and $0.40$.
\label{ptilde_pg} }
\end{center}
\end{figure}

As in Section \ref{eva}, we now contrast with the quenched theory,
where it is well-known that the symmetry is broken for sufficiently
large values of $b$.  The results are summarized in Fig.~\ref{ptilde_pg}.
Comparing to Fig.~\ref{ptilde_m0001}, what one sees is that there is
a significant difference in behavior between $b=0.10, 0.20$ versus $b=0.30, 0.40$
in the quenched case, which did not occur in the unquenched case.
The latter two values of $b$ are significantly lower and actually trend somewhat
downward as $N$ is increased.  
The fits to \myref{ffque} are shown in Table \ref{fqutab}; the
$b=0.40$ case does not give a good fit, but it is clear that the
asymptotic value in the large $N$ limit is significantly smaller
than in the unquenched case.  Setting aside the poor quality of
the fit in this case, the large $N$ limit of $A=0.2939(38)$ for
$b=0.40$ is far below the value in the highly symmetric case of $b=0.10$,
$A=0.50074(30)$ which is essentially $1/2$.  This is indicative of the strong breaking of
center symmetry for this large value of $b=0.4$, as usual in the large $N$ limit.

\begin{table}
\begin{center}
\begin{tabular}{ccccc}
\hline
b & A & B & C & $\chi^2/\text{d.o.f.}$ \\
\hline
0.10 & 0.50074(30) & -1.177(22) & 2.12(22) & 1.53 \\
0.20 & 0.4425(13)  &  -1.147(66) & 2.55(57) & 1.75 \\
0.30 & 0.3511(17) & 0.175(96) & -4.40(81) & 2.20  \\
0.40 & 0.2939(38) & 0.91(22) & -7.8(1.8)  & 4.21 \\
\hline
\end{tabular}
\caption{Results for the fit of the quenched data to Eq.~\myref{ffque}.
\label{fqutab}}
\end{center}
\end{table}

\section{Spontaneous symmetry breaking for the single site lattice}
\label{spontaneous}
There is a free energy of the (traced) Polyakov loop $P$, $F_N(P)$, for each value of $N$.  
Spontaneous symmetry breaking would mean that there is a very large
barrier between the $N$ different minima of $F_N(P)$, because otherwise
the finite tunneling probability would destroy any order.  In fact, this
is the reason why the thermodynamic limit must be taken in order
to have true spontaneous symmetry breaking, because we need a very large
number of degrees of freedom in order to produce the 
corresponding large barriers.  These barriers
can only arise in the thermodynamic limit, which is formally $N \to \infty$ in the
single site lattice theory.  It follows that the vacua are quite close to each other
in the relevant limit, so the vanishingly small tunneling between them is a subtle issue.
It is necessary for the distribution of $\arg (P)$ to become highly nonuniform
in the thermodynamic limit.  The point is that on the single
site lattice there is a free
energy density, $f_N(P) = F_N(P)/N^2$, which is a reasonable function, but
\beq
\exp(-F_N(P)) = \exp(-N^2 f_N(P))
\eeq
gets vanishingly small weight at all but the minima of $f_N(P)$ in the large $N$ limit.
Any external perturbation will then freeze it into one of those minima, and the
barrier for tunneling is effectively infinite.

This discussion makes it clear that the physics of spontaneous symmetry breaking
in the single site lattice theory must be understood in the large $N$ limit.
It is for this reason that we have studied a method that allows for an $N \to \infty$
extrapolation in the previous section.  Certainly a vanishing ratio
\myref{2parR} or \myref{3parR}
would be a clear indication of a uniform distribution, and hence no spontaneous
symmetry breaking.  However, from our understanding of spontaneous symmetry breaking
as arising from infinite barriers, we see that small but finite ratio (such as
was found for $b=0.10$ in the $m=0.1$ case) does not
indicate a broken symmetry phase, since there is still a finite density of
states in the tunneling region that lies between the vacua.  Indeed, the
ratio should become rather large if the symmetry is spontaneously broken in the
$N \to \infty$ limit.  Correspondingly,
a fingered traced Polyakov loop distribution is not
a clear indicator of a broken phase, since there still is a nonzero density
of states in the central tunneling region.  Only in the case where multiple
observables indicate a high degree of nonuniformity in the thermodynamic
limit can one conclude with any confidence that center symmetry is spontaneously
broken.  Examples of this are the unquenched theory with $b=0.30$ and $m=0.1$,
or the quenched theory with $b=0.30, 0.40$.

\section{Conclusions}
We have found that on the single site lattice
with Ginsparg-Wilson-type fermions (in our case M\"obius),
center symmetry is unbroken in the large $N$ limit if the fermions are
effectively massless.  On the other hand, at the relatively
large mass value of $m=0.1$, we find that sufficiently large $b$
will induce spontaneous center symmetry breaking at large $N$.
This agrees with the fact that in the continuum large $N$ limit the
one loop effective potential for the eigenvalues
of the (untraced) Polyakov loop shows that they
repel and are uniformly distributed for any $N_f \not= 0$ \cite{Kovtun:2007py},
provided the fermions are massless.

A different result has been obtained recently using
an approach which truncates the link matrices to diagonal matrices
only consisting of their eigenvalues, finding that in the large $N$ limit, $N_f=1$
has spontaneously broken center symmetry \cite{Lohmayer:2013spa}.
It may be that this truncation somehow misses important physics.
Our nonperturbative lattice results agree with Ref.~\cite{Hietanen:2012ma}
which found that to a good approximation \myref{plmeas} was $1/2$,
corresponding to Polyakov loops near zero on most configurations,
and hence unbroken center symmetry.  Similarly, the older study 
\cite{Bringoltz:2009kb} which used Wilson fermions found unbroken
center symmetry for a wide range of fermion masses.  We believe that
our present study goes beyond these results by fitting the eigenvalue
distribution to a function of $N$ so that the large $N$ limit can
be taken.  Since true spontaneous symmetry breaking can only be
obtained in this limit, we would argue that it is important to
develop a quantitative method that is amenable to such an extrapolation.
We have also presented an argument that simply having a nonuniform
distribution does not necessarily imple spontaneous symmetry breaking,
since one must consider the possibility of tunneling between the
$N$ ground states in the large $N$ limit.  By comparing to the
quenched case, where the fate of center symmetry is understood,
we have concluded that the nonuniformity must be rather large.

Our simulation results also agree with
the nonperturbative findings of \cite{Hietanen:2009ex,Hietanen:2010fx}, 
where the theory of a single Majorana fermion
(``half a Dirac flavor'')
in the adjoint representation was studied on a single site lattice.
For $N=11$ and $b=7$ they found in \cite{Hietanen:2009ex} that the quantity
\myref{plmeas} was approximately $1/2$ for the right range of the
Wilson kernel mass (what we are calling $m_5$).
In \cite{Hietanen:2010fx} they found that for $N = 11, 15$ and $18$ the quantity \myref{plmeas} was 
approximately $1/2$ for $b=5$, provided
the fermion mass $m$ in lattice units satisfied $m < 0.1$.

One direction for further research is to increase the values of $b$ that
we are able to probe.  This would require abandoning the Metropolis
algorithm in favor of something like the rational hybrid Monte Carlo
algorithm.

\section*{Acknowledgements}
The authors were supported in part by the Department of Energy,
Office of Science, Office of High Energy Physics.  Both authors
received support from Grant No.~DE-FG02-08ER41575 and JG was 
supported in part by Grant No.~DE-SC0013496.  We thank R.~Narayanan
for helpful comments.  We are particularly indebted to a referee
for numerous suggestions that led to further studies and improvements
to this article.

\bibliography{ekcs}
\bibliographystyle{apsrev4-1}

\end{document}